\documentclass[12pt,a4paper]{article}

\usepackage{graphicx}

\usepackage{cite}

\makeatletter
\renewcommand{\@biblabel}[1]{}
\makeatother

\begin{document}

\setlength{\baselineskip}{0.77cm}

\newcommand{\EQ}{Eq.~}
\newcommand{\EQS}{Eqs.~}
\newcommand{\FIG}{Fig.~}
\newcommand{\FIGS}{Figs.~}
\newcommand{\SEC}{Sec.~}
\newcommand{\SECS}{Secs.~}

\begin{center}
{\Large \bf VIP-club phenomenon: emergence of elites and 
masterminds
in social networks}
\bigskip

{\Large Naoki Masuda${}^{1}$ and
Norio Konno${}^{2}$}

\bigskip

${}^1$ Laboratory for Mathematical Neuroscience,
RIKEN Brain Science Institute, 2-1, Hirosawa, Wako, Saitama, 351-0198,
Japan\\
${}^2$ Faculty of Engineering,
Yokohama National University,
79-5, Tokiwadai, Hodogaya, Yokohama, 240-8501, Japan
\end{center}



\section*{Abstract}
Hubs, or vertices with large degrees, play massive roles in, for
example, epidemic dynamics, innovation diffusion, and synchronization
on networks.  However, costs of owning edges can motivate agents to
decrease their degrees and avoid becoming hubs, whereas they would
somehow like to keep access to a major part of the network.  By
analyzing a model and tennis players' partnership networks, we show
that combination of vertex fitness and homophily yields a VIP club
made of elite vertices that are influential but not easily accessed
from the majority. Intentionally formed VIP members can even serve as
masterminds, which manipulate hubs to control the entire network
without exposing themselves to a large mass. From
conventional viewpoints based on network topology and edge
direction, elites are not distinguished from many other vertices.
Understanding
network data is far from sufficient; individualistic factors greatly
affect network structure and functions per se.

\bigskip

{\small
corresponding author: Naoki Masuda,
Laboratory for Mathematical Neuroscience,
RIKEN Brain Science Institute, 2-1, Hirosawa, Wako, Saitama, 351-0198,
Japan,  email:masuda@brain.riken.jp, TEL:+81-48-467-9664, FAX:+81-48-467-9693.
We thank Etsuro Segawa for helpful discussion.
This work is supported by Special
Postdoctoral Researchers Program of RIKEN.}

\newpage

\section{Introduction}\label{sec:introduction}

Real networks are neither regular nor completely random. They
are random to some extent and equipped with short diameters and the
clustering property \cite{SW}. Many networks are also scale-free.
In other words,
the number of edges per vertex, which is denoted by $k$, follows the
power-law distribution: $p(k)\propto k^{-\gamma}$
($\gamma>0$). This implies the 
existence of a considerable number of hubs, or 
vertices with huge degrees,
beyond what is expected of a stereotypical bell-shaped $p(k)$ such as
the Gaussian distribution.
Hubs play central roles in, for
example, robustness against random and intentional attacks against 
vertices or edges,
propagating innovations,
disease spreading, and synchronizing dynamical agents placed on vertices
\cite{Albert02,Newman03SIAM,Pastor}. Naturally,
real data analysis has focused on
finding and characterizing hubs, including the evaluation of $\gamma$
\cite{Albert02,Newman03SIAM,ZhouMondragon}.

However, vertices other than hubs are also important on other
occasions. As an example, let us imagine a one-shot transmission
of innovation
or rumor starting from an arbitrarily chosen 
vertex.  Diffusion studies
establish that very first adopters are regarded too radical or
immature to directly communicate the new information to a
majority. Instead, so-called early adopters or opinion leaders,
presumably with high social statuses, credibility, or high
connectivity as represented by hubs, receive the new information
from the first adopters
and boost its propagation \cite{Rogers}.  Apart from the obvious role
of hubs, however, properties of the first adopters determine the
dominant timescale and even the success probability of diffusion,
which are important in applications such as marketing and epidemics.
In infinite particle systems used in physics and mathematics,
such as the percolation and the contact
process, the role of hubs (resp. early adopters) are manifested
when there are initially many (resp. few) vertices.

If connectivity is flexible,
agents can even be motivated to avoid becoming hubs
and take advantage of other hubs. For example, computer
viruses on scale-free networks proliferate by spreading through hubs
since they are more accessible from others
\cite{Pastor}. Then, it is better for the system cracker to hide
behind hubs and exploit them, than to expose themselves to a major
part of networks as hubs, which raises a risk to be detected by the
authority or other vertices.  
Similarly, intention of manipulating hubs may be present in economical
behavior, 
politics, and marketing.  The tradeoff between a cost
of directly spanning edges and a benefit of having direct and indirect
access to others can be formalized by a utility function such as
\begin{equation}
\sum_{l=1}^{\infty} k_l \delta^l - C k,
\label{eq:utility}
\end{equation}
where $k_l$ is the number of vertices at distance $l$ from a reference
vertex, $k\equiv k_1$, $0<\delta<1$ is a discount factor
\cite{Jackson,Bala,AlisonWatts}, and $C k$ ($C>0$) is the cost of
maintaining edges or being exposed to others. 
Equation~(\ref{eq:utility}) is also related to
the growing scale-free network model, in which new
vertices with a typically small constant $k$
 are consecutively added to a network.
Based on the assumption of the preferential attachment,
newcomers attempt to get linked to
hubs \cite{Albert02,Newman03SIAM}.
If
edges are dynamically created and removed according to
\EQ(\ref{eq:utility}) or similar utility functions, networks
typically end up with wheels, stars, or complete networks
\cite{Jackson,Bala,AlisonWatts}.
These mathematical results imply realistic network structures as a
result of evolution; for example, stars indicate hubs.  However, real
networks seem to deviate from wheels, stars and complete graphs that
the utility model, which usually accompanies strong constraints,
predicts. We call a vertex with a large utility value an elite or a 
mastermind.

In this paper, we
explore how hubs
and elites emerge, function, and interact, which has been
neglected except in a few studies \cite{Lau,Anghel}.
Based on thresholding and homophily explained in \SEC\ref{sec:thresh_homo},
we propose a network model in \SEC\ref{sec:model}. In
\SEC\ref{sec:vip}, we show
that combination of thresholding and homophily naturally
generates elites in networks.
We also analyze tennis tournament data in \SEC\ref{sec:tennis}.


\section{Thresholding and Homophily}\label{sec:thresh_homo}

Let us introduce the intrinsic weight
of the $i$th vertex
denoted by $w_i$. It quantifies the potential to
win edges, such as physical ability, fame, and social status
\cite{Bianconi,Caldarelli,Goh01PRL,Boguna,Masuda_THRESH,Barrat,Masuda_SPTH}.
Depending on situations, the direction of influence from vertices with
larger weights to ones with smaller weights can be deduced
independently of the predefined edge direction \cite{Anghel}.  For
instance, whether a computer virus at a host can invade another
depends on the relative security level of hosts, or vertices.
Networks with weight-driven edge direction also underly human
relationships.

Another key element is homophily, which means that similar agents,
particularly humans, tend to flock together. Many real data from
individual questionnaires \cite{Marsden,Mcpherson01,Mcpherson87},
diffusion studies \cite{Rogers}, and analysis of online communities
\cite{Adamic03SOC,Adamic03} support homophily according to nominal (e.g.  race,
hobbies, sex, religious preference, personal traits) and graduated
(e.g. physical distance, age, education, social
status) parameters. Homophily in terms of vertex degrees, or rates of social
contacts, underlies the fact that the degrees of adjacent vertices
are
positively correlated in social networks \cite{Newman03SIAM}.
Various models of real networks and social interactions incorporate
homophily. For
example, Granovetter (1973) proposed that one is mainly connected to
similar others, and some weak ties also exist to bridge heterophilous
individuals. Then, the small-world property results because weak ties
shorten the network diameter and abundant homophilous connectivity
enhances clustering \cite{SW}. Hierarchical networks in which vertices
with closer hierarchical levels are more likely to be adjacent are
used to address search and congestion problems on networks
\cite{Watts02SCI,Dodds}.  We note that hierarchy is also reminiscent
of weight-driven edge direction mentioned before.
Other examples include the
cultural exchange models with general homophily \cite{Axelrod97} and
the gravity models with spatial homophily
\cite{Zipf,Barrat,Masuda_SPTH}.  We focus on homophily based on 
graduated parameters because
vertices are equipped with graduated intrinsic weights.

\section{Model}\label{sec:model}

We show in the context of scale-free networks
that combination of thresholding and homophily yields networks with
elites.  Let us prepare $n$
vertices and choose $w_i$ ($1\le i\le n$) randomly and independently
from a distribution $f(w)$. We start with the threshold graph, in
which two vertices with weights $w$ and
$w^{\prime}$ are connected if $w+w^{\prime}\ge \theta$
\cite{Caldarelli,Boguna,Masuda_THRESH}. A larger $w$
induces a larger vertex degree $k$, which is the so-called rich-club
phenomenon \cite{ZhouMondragon}. Scale-free networks with the
small-world properties actually result from various $f(w)$;
$p(k)\propto k^{-2}$ from the exponential distribution $f(w)=\lambda
e^{-\lambda w}$ ($w\ge 0$) \cite{Caldarelli,Boguna,Masuda_THRESH} and
$p(k)\propto k^{-(a+1)/a}$ from the Pareto distribution $f(w)\propto
w^{-a-1}$ ($w\ge w_0$, $\exists w_0>0$)
\cite{Masuda_THRESH}.  We now supply a homophily rule by
making the connection probability decreasing in $|w^{\prime}-w|$.  For
simplicity, an edge is assumed to be created only when $|w^{\prime}-w|\le
c$. Together with $w+w^{\prime}\ge\theta$, a vertex with $w$ is
adjacent to vertices whose weights satisfy
\begin{equation}
w^{\prime}\in \left\{
\begin{array}{lll} 
&\emptyset, & (w<\frac{\theta-c}{2})\\
&[\theta-w, w+c], & (\frac{\theta-c}{2}\le w<\frac{\theta+c}{2})\\
&[w-c, w+c]. & (w\ge\frac{\theta+c}{2})
\end{array} \right.
\label{eq:thresh1}
\end{equation}
In the limit $c\to\infty$,
\EQ(\ref{eq:thresh1}) becomes
$\theta-w \le w^{\prime} < \infty$, returning to 
the original threshold graph.
We obtain $k$ as a function of
$w$ by integrating $f(w^{\prime})$ over the
range given in \EQ(\ref{eq:thresh1}).  If $f(w)$ is monotone
decreasing for
$w>w_c\equiv (\theta+c)/2$, $k(w)$ is maximized at
$w=w_c$.  Even if not,
sufficiently large $w$ with
\begin{equation}
1-F(w-c) < F\left( \frac{\theta+3c}{2}\right) 
- F\left( \frac{\theta-c}{2}\right)
\end{equation}
satisfies $k(w)<k\left(\left(\theta+c\right)/2\right)$, meaning that
$k(w)$ takes the maximum at $w=w_c\in (\frac{\theta+c}{2},\infty)$.
In both cases, vertices with $w\cong w_c$ are hubs. Elites are
vertices with $w\gg w_c$ and not exposed via direct edges to the major
group of vertices with small $w$.

The psychological Weber-Fechner law dictates that
vertices may sense relative rather than absolute differences
in weights.
To mimic this, let us modify the homophily condition to
$|w^{\prime}-w|/(w+w^{\prime})<c$ ($c<1$).
Then, the counterpart of \EQ(\ref{eq:thresh1}) reads
\begin{equation}
w^{\prime}\in \left\{
\begin{array}{lll} 
&\emptyset, & (w<\frac{1-c}{2}\theta)\\
&[\theta-w, \frac{1+c}{1-c}w], &
(\frac{1-c}{2}\theta\le w<\frac{1+c}{2}\theta)\\
&[\frac{1-c}{1+c}w, \frac{1+c}{1-c}w].
& (w\ge\frac{1+c}{2}\theta)
\end{array} \right.
\label{eq:thresh_Fechner}
\end{equation}
For $k(w)$ to be maximized at
$w=(1+c)\theta/2$, 
$f(w)$ fulfilling
\begin{equation}
\left(\frac{1+c}{1-c}\right)^2 f\left( \left(\frac{1+c}{1-c}\right)^2
w \right) < f(w) \quad \left(w> \left(1-c\right)/2\theta\right)
\label{eq:peak_Fechner}
\end{equation}
is required. However, as in the previous case, the maximum of $k(w)$ 
at $w = w_c > (1+c)\theta/2$
is assured even if \EQ(\ref{eq:peak_Fechner})
is violated.

To consider weight-driven edge direction,
we again apply $|w^{\prime}-w|\le c$. Since
a directed edge $w\to w^{\prime}$ may form only when
$w>w^{\prime}$, a vertex with $w$
sends directed edges to ones with
\begin{equation}
w^{\prime}\in \left\{
\begin{array}{lll} 
&\emptyset, & (w<\frac{\theta}{2})\\
&[\theta-w, w], & (\frac{\theta}{2}\le w<\frac{\theta+c}{2})\\
&[w-c, w], & (w\ge\frac{\theta+c}{2})
\end{array} \right.
\label{eq:thresh1_directed}
\end{equation}
which is essentially the same as \EQ(\ref{eq:thresh1}).

\section{VIP-club Phenomenon}\label{sec:vip}

\begin{figure}
\begin{center}
\includegraphics[height=1.75in,width=2.25in]{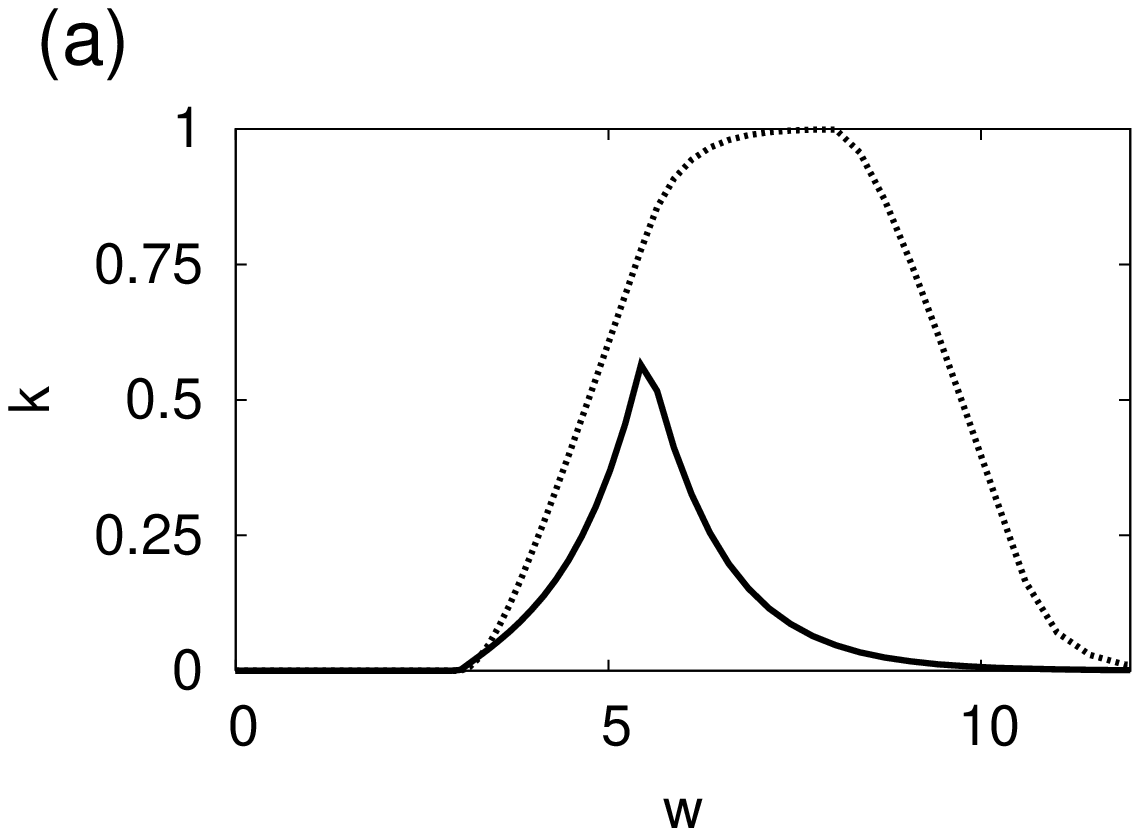}
\includegraphics[height=1.75in,width=2.25in]{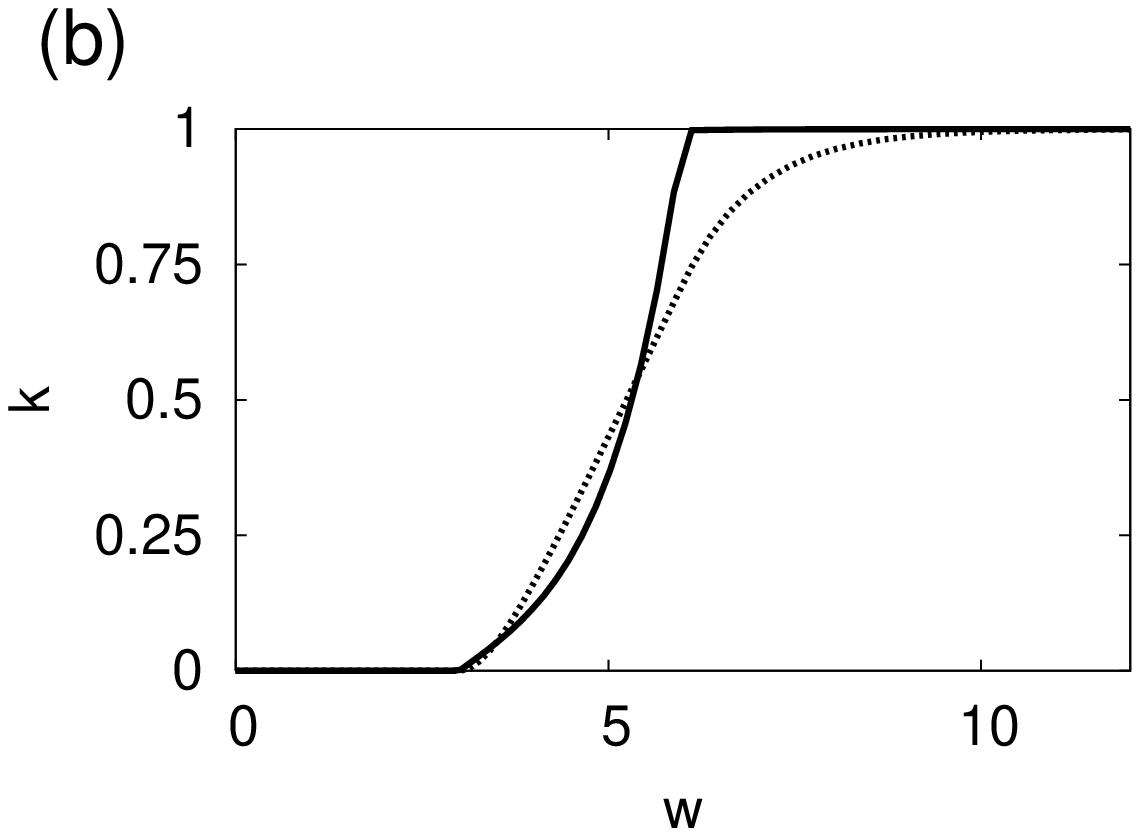}
\includegraphics[height=1.75in,width=2.25in]{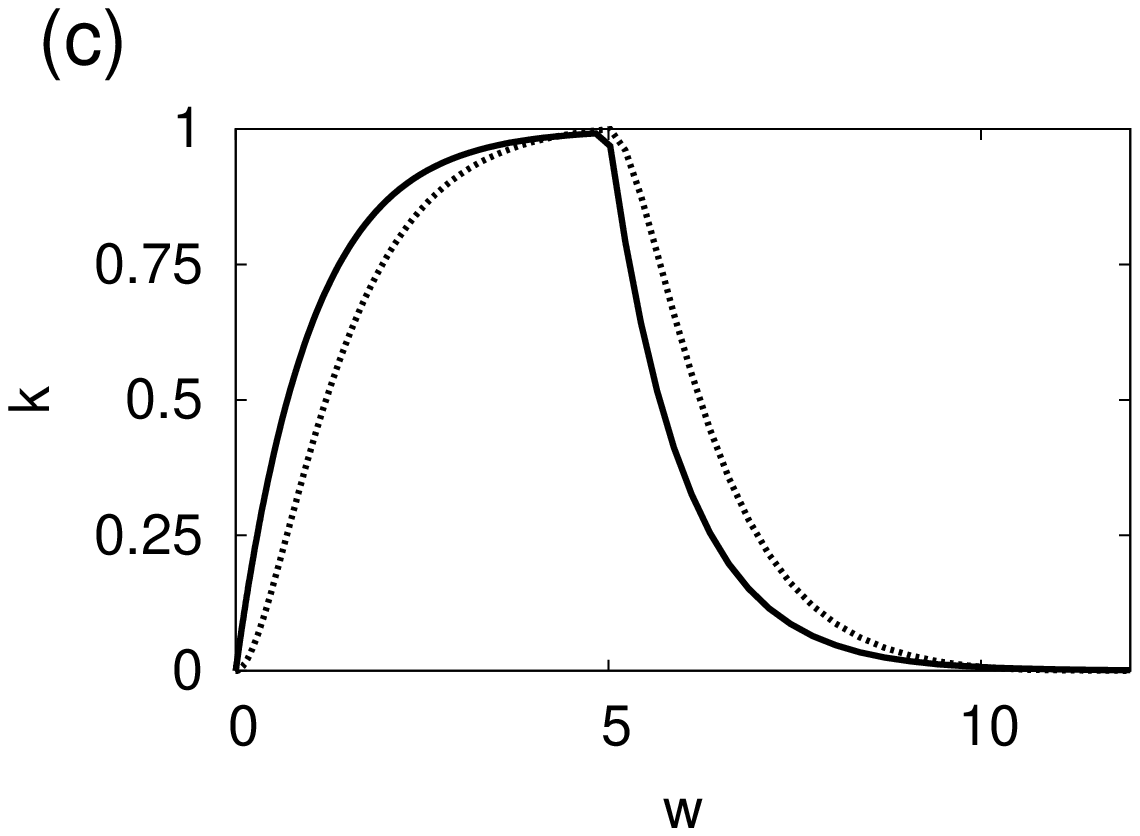}
\includegraphics[height=1.75in,width=2.25in]{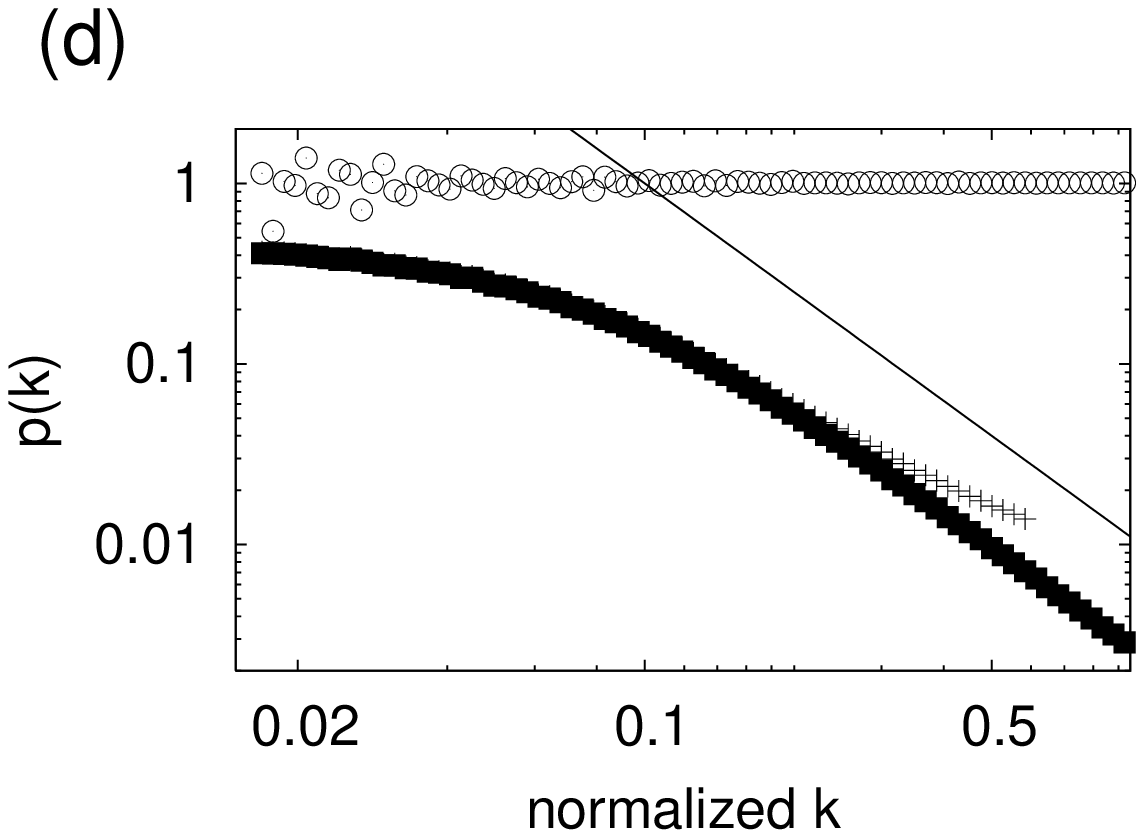}
\caption{}
\label{fig:exp_th}
\end{center}
\end{figure}

For concreteness, we set
$f(w)=\lambda e^{-\lambda w}\; (w\ge 0)$.
The following results hold
as long as $f(w)$ largely decreases when $w\ge w_c$,
which is supported by real data
\cite{Zipf,Masuda_THRESH,Masuda_SPTH}.
Based on \EQ(\ref{eq:thresh1_directed}),
the vertex degree as a function of
the weight is represented by
\begin{equation}
k(w) = \left\{ \begin{array}{ll}
0, & (w<\frac{\theta}{2})\\
e^{-\lambda(\theta-w)} - e^{-\lambda w}, &
(\frac{\theta}{2}\le w<\frac{\theta+c}{2})\\
e^{-\lambda(w-c)} - e^{-\lambda w},
& (w\ge\frac{\theta+c}{2})
\end{array}\right.
\label{eq:k(w)_thresh_master1}
\end{equation}
for $\theta\ge c$, and 
\begin{equation}
k(w) = \left\{ \begin{array}{ll}
0, & (w<\frac{\theta}{2})\\
e^{-\lambda(\theta-w)} - e^{-\lambda w}, &
(\frac{\theta}{2}\le w<\theta)\\
1 - e^{-\lambda w}, & (\theta\le w< c) \\
e^{-\lambda (w-c)} - e^{-\lambda w}, & (w\ge c)
\end{array}\right.
\label{eq:k(w)_thresh_master2}
\end{equation}
for $\theta < c$. 
As shown in \FIG\ref{fig:exp_th}(a) for 
$(\theta, c) = (6, 5)$ [\EQ(\ref{eq:k(w)_thresh_master1})],
$k(w)$ (solid line) has a single peak.
Similar upshots result from
\EQ(\ref{eq:thresh1}), (\ref{eq:thresh_Fechner}), or
(\ref{eq:k(w)_thresh_master2}) if the thresholding
and the homophily are roughly balanced.

With a utility function like \EQ(\ref{eq:utility}) in mind,
we derive $k_2(w)$, which is
the number of the vertices within two hops from a vertex with
weight $w$. Although we exclude the reference vertex itself, this
subtlety does not matter when the network is large enough.
The neighbor's weight $w^{\prime}$ satisfies 
\EQ(\ref{eq:thresh1_directed}). The weight of the neighbor's neighbor,
which is denoted by $w^{\prime\prime}$, similarly satisfies
\begin{equation}
w^{\prime\prime}\in \left\{
\begin{array}{lll} 
&\emptyset, & (w^{\prime}<\frac{\theta}{2})\\
&[\theta-w^{\prime}, w^{\prime}],
 & (\frac{\theta}{2}\le w^{\prime}<\frac{\theta+c}{2})\\
&[w^{\prime}-c, w^{\prime}]. & (w^{\prime}\ge\frac{\theta+c}{2})
\end{array} \right.
\label{eq:thresh1_directed2}
\end{equation}
For a given $w$, we integrate the density of vertices
$f(w^{\prime\prime})$ over the range
compatible with \EQS(\ref{eq:thresh1_directed})
and (\ref{eq:thresh1_directed2}) to obtain
\begin{equation}
k_2(w) \propto \left\{ \begin{array}{ll} 
0, & (w<\frac{\theta}{2})\\
\lambda e^{-\lambda\theta}\left( w - \frac{\theta}{2}\right)
+ \frac{e^{-2\lambda w}-e^{-\lambda\theta}}{2},
 & (\frac{\theta}{2}\le w<\frac{\theta+c}{2})\\
\frac{\lambda c e^{-\lambda\theta} 
-(e^{\lambda c}-1)e^{-2\lambda w}}{2}, & 
(\frac{\theta+c}{2}\le w<\frac{\theta+2c}{2})\\
\frac{\lambda(\theta-2w+3c)+1}{2} e^{-\lambda\theta}
+ \frac{e^{-2\lambda w}(-e^{2\lambda c}-e^{\lambda c} + 1)}{2},
& 
(\frac{\theta+2c}{2}\le w<\frac{\theta+3c}{2})\\
\frac{(e^{\lambda c}-1)(e^{2\lambda c}-1)}{2} e^{-2\lambda w},
& (w\ge\frac{\theta+3c}{2})
\end{array} \right.
\label{eq:thresh2_directed}
\end{equation}
for $\theta\ge c$, and 
\begin{equation}
k_2(w) \propto \left\{ \begin{array}{ll} 
0, & (w<\frac{\theta}{2})\\
\lambda e^{-\lambda\theta}\left( w - \frac{\theta}{2}\right)
+ \frac{e^{-2\lambda w}-e^{-\lambda\theta}}{2},
 & (\frac{\theta}{2}\le w<\theta)\\
\frac{\lambda \theta+1}{2} e^{-\lambda\theta} 
- e^{-\lambda w} - \frac{e^{-2\lambda w}}{2}, & 
(\theta\le w< c)\\
\frac{\lambda \theta+1}{2} e^{-\lambda\theta} +
\frac{-e^{-\lambda c}+e^{-2\lambda w}-e^{(-2w+c)\lambda}}{2}, & 
(c\le w< c+\frac{\theta}{2})\\
\left(1-\lambda\left(w-c-\theta\right)\right)e^{-\lambda\theta}
+\frac{1}{2}e^{-2\lambda w}\left(
-e^{2\lambda c}-e^{\lambda c}+1\right)
-\frac{1}{2}e^{-\lambda c}, & 
(c+\frac{\theta}{2}\le w< c+\theta)\\
\frac{1}{2} e^{-2\lambda w}\left( -e^{2\lambda c} - 
e^{-\lambda c} +  1\right) -\frac{1}{2}e^{-\lambda c}
+ e^{-\lambda (w-c)}, & 
(c+\theta\le w< 2c)\\
\frac{(e^{\lambda c}-1)(e^{2\lambda c}-1)}{2} e^{-2\lambda w},
& (w \ge 2c)
\end{array} \right.
\label{eq:thresh3_directed}
\end{equation}
for $\theta < c$. Figure~\ref{fig:exp_th}(a) shows that $k_2(w)$ (dotted
line) has a unique peak at $w = \overline{w}_c>w_c$ and that $k_2(w)$
decays more slowly in $w$ than $k(w)$ (solid line) does.  Under
\EQ(\ref{eq:utility}) or a simpler utility function with only two
terms involving $k$ and $k_2$ \cite{Jackson}, vertices with
$w\cong \overline{w}_c$ are elites or
masterminds, whereas vertices with $w\cong w_c$ are hubs. The
masterminds are linked to the majority of vertices with small $w$ and
presumably large $f(w)$ only indirectly via hubs. Owing to homophily, they
flock together with others
with $w\cong \overline{w}_c$, which are usually rare.  We call
it VIP-club phenomenon in contrast to the rich-club phenomenon
\cite{ZhouMondragon}, in which larger $w$ simply means larger $k(w)$.

Figure~\ref{fig:exp_th}(b) shows
$k(w)$ (solid line) and $k_2(w)$ (dotted line)
for $(\theta, c) = (6, 100)$ with which
homophily is practically absent.
In accordance with the standard
threshold model, $k(w)$ increases monotonically in $w$,
and vertices with large $w$ serve as hubs
\cite{Caldarelli,Boguna,Masuda_THRESH}. This is the rich-club but
not VIP-club phenomenon.

With homophily only (\FIG\ref{fig:exp_th}(c),
$(\theta, c) = (0, 5)$), the majority vertices, which have small $w$,
own large degrees because
homophily simply interconnects these vertices in the absence of
thresholding.
What
reduces $k(w)$ for small $w$ in \FIG\ref{fig:exp_th}(c) is the lower
bound of the exponential weight distribution at $w=0$, which is
nonessential. In consequence, hubs abound in the network, and the
VIP-club does not form.  Moreover, the homophily-only configuration
is unrealistic for two reasons. First, 
$p(k)$ becomes flat
(circles in \FIG\ref{fig:exp_th}(d)),
which contradicts real data
\cite{SW,Albert02,Newman03SIAM}. 
In contrast, 
although homophily prohibits huge hubs,  $p(k)$ of a network with both
homophily and thresholding is scale-free
(crosses), as is $p(k)$ of the 
the threshold graph
without homophily (squares).
This is because elites are
scarce even if they exist.  Second, too strong homophily mars
communication between vertices with distant weights.
With the density of edges given,
the diameter becomes too large in a strongly
homophilous network due to the
scarceness of shortcuts bridging heterophilous
vertices \cite{Granovetter73,SW,Rogers,Adamic03}.
The thresholding effect counteracts the homophily effect to
render a network small-world.  In sum, the VIP-club phenomenon
requires both homophily and thresholding in our framework.



\begin{figure}
\begin{center}
\includegraphics[height=1.75in,width=2.25in]{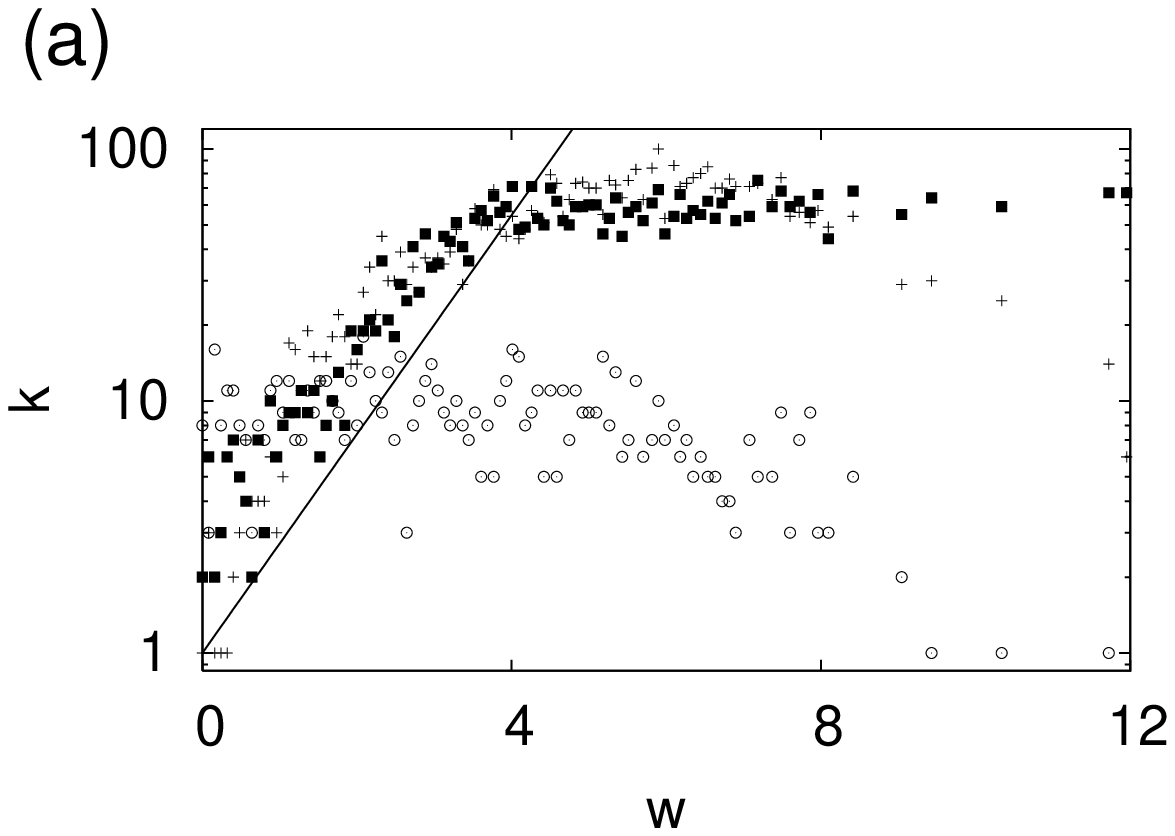}
\includegraphics[height=1.75in,width=2.25in]{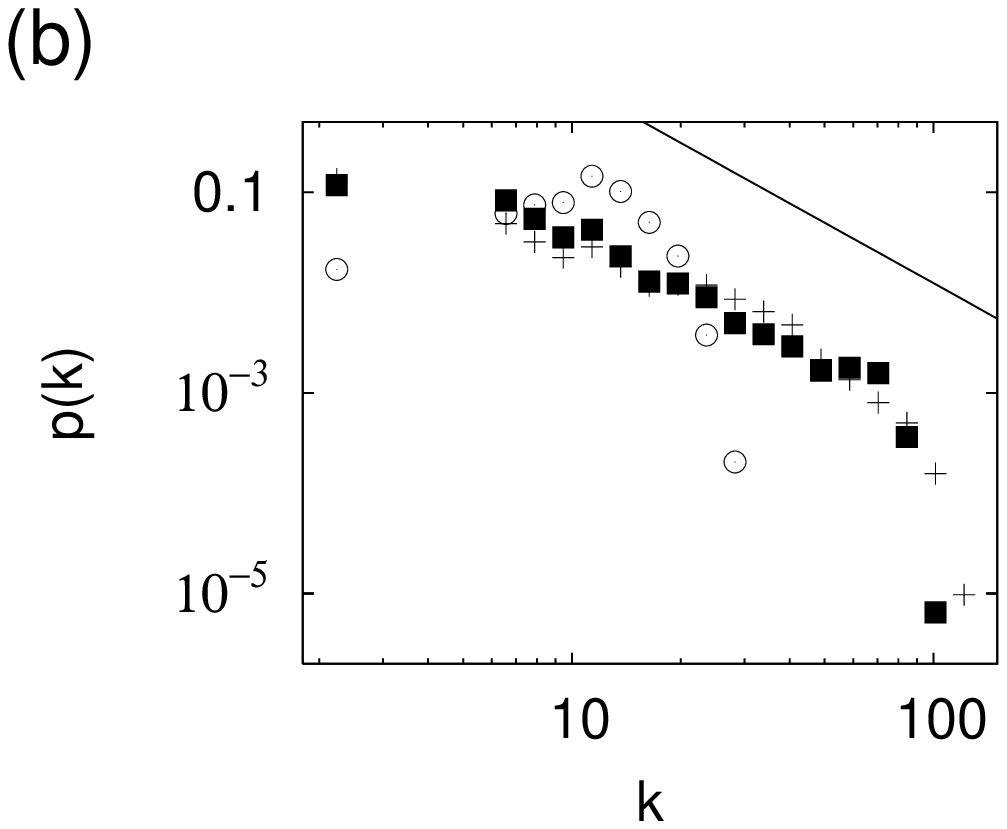}
\caption{}
\label{fig:exp}
\end{center}
\end{figure}

Since the analysis developed so far
%
%
is for deterministic dense
networks, let us numerically examine sparse networks with
stochasticity.  An edge is assumed to
form between vertices with weights $w$ and
$w^{\prime}$ with probability proportional to
$e^{-\beta_2|w^{\prime}-w|} /(1+e^{-\beta_1(w+w^{\prime}-\theta)})$
where $\beta_1$ and $\beta_2$ are the inverse temperatures.  We set
$n=50000$, $\lambda=1$, $\theta=6$, and the mean degree 10.
Figures~\ref{fig:exp}(a) and \ref{fig:exp}(b) show $k(w)$ and $p(k)$,
respectively, for $(\beta_1, \beta_2)=$ $(1.5, 0.5)$ (both
homophily and
thresholding, plotted by crosses), $(1.5,0)$ (thresholding only,
squares), and $(0,0.5)$ (homophily only, circles).  These results are
roughly consistent with \FIG\ref{fig:exp_th}.

\begin{figure}
\begin{center}
\includegraphics[height=1.75in,width=2.25in]{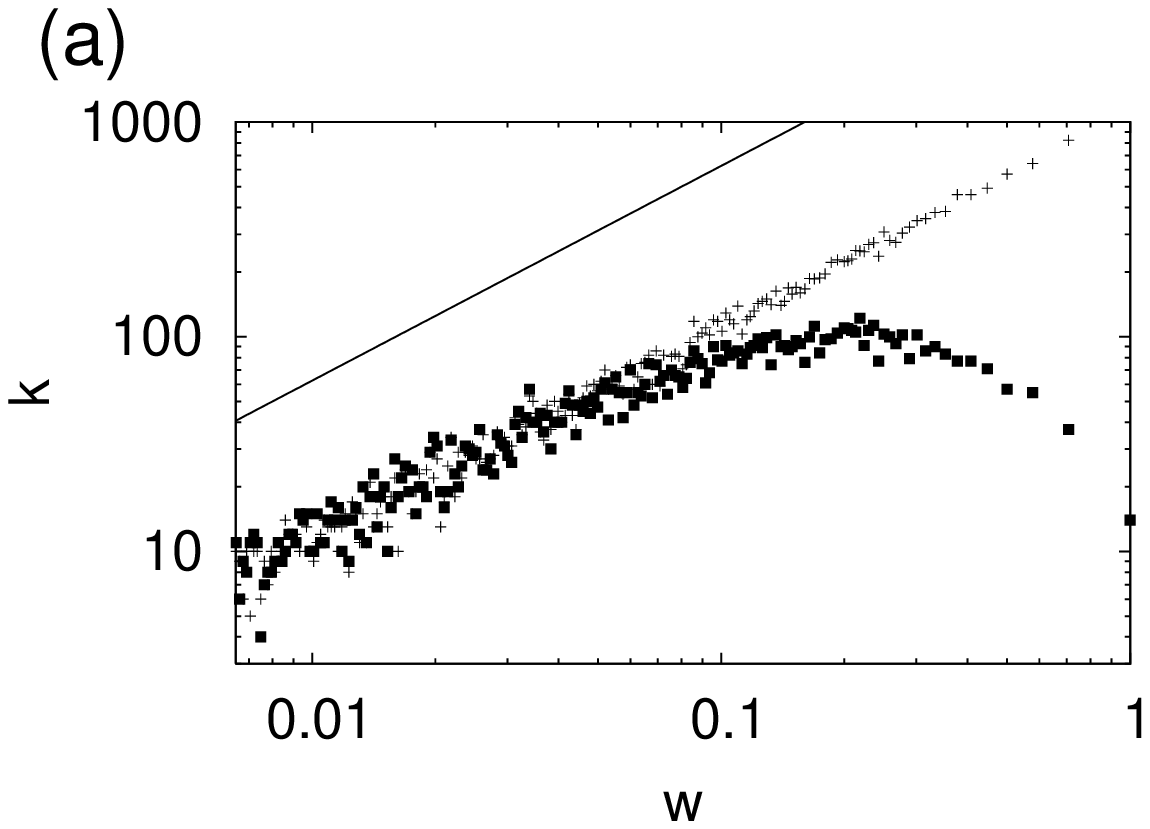}
\includegraphics[height=1.75in,width=2.25in]{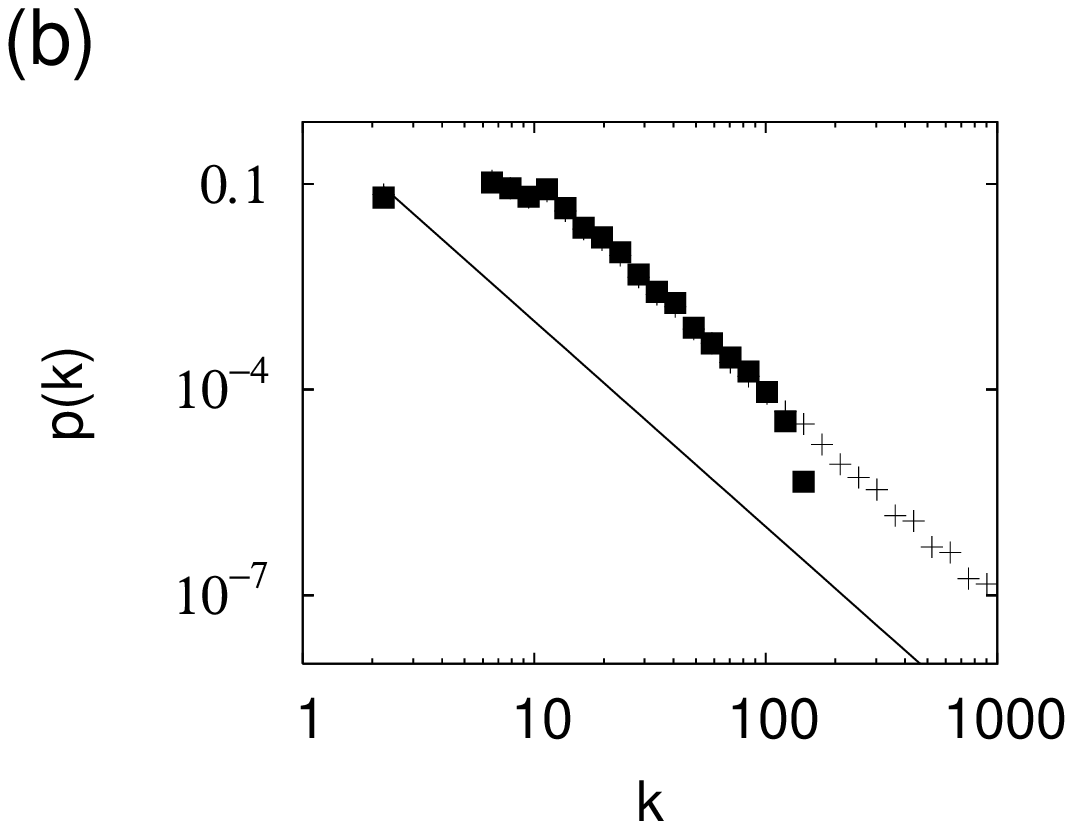}
\caption{}
\label{fig:goh}
\end{center}
\end{figure}

Let us next examine a scale-free network in which $w_i = i^{-\alpha}$
is assigned to the $i$th vertex \cite{Goh01PRL}. Then, a pair of
vertices is picked according to the distribution $p(i) =
w_i/\sum^n_{i=1} w_i$, and an edge is created if they are not yet
adjacent. This procedure is repeated until the mean degree 10 is
reached. As a result, we obtain $p(k)\propto k^{-\gamma}$ with
$\gamma=(1+\alpha)/\alpha$ \cite{Goh01PRL}. A type of thresholding is
embedded in this algorithm. We implement homophily by supposing that
an edge forms with probability $e^{-\beta_2|w^{\prime}-w|}$ after two
vertices with weights $w$ and $w^{\prime}$ are selected. The results
shown in \FIG\ref{fig:goh} for $n=50000$ and $\alpha=0.5$ with
homophily present ($\beta_2 =5$, crosses) and absent ($\beta_2=0$,
squares) are consistent with \FIGS\ref{fig:exp_th} and
\ref{fig:exp}. Although we have presented just two examples, our
framework is applicable to other networks with vertex fitness.

\section{Analysis of Tennis Players' Networks}\label{sec:tennis}

\begin{figure}
\begin{center}
\includegraphics[height=1.75in,width=2.25in]{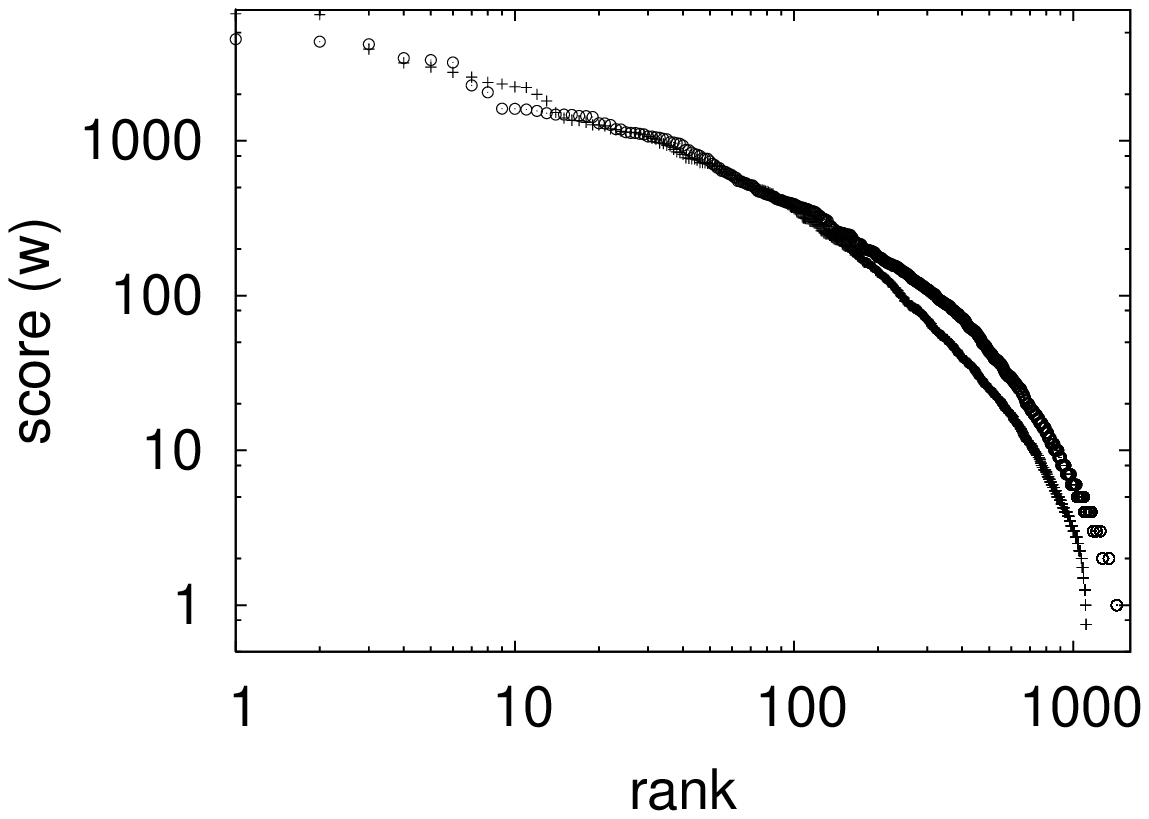}
\caption{}
\label{fig:rank-score}
\end{center}
\end{figure}

In social networks with homophily, graduated weight variables often
exist but are difficult to measure. In
the professional tennis community,
players are ranked based on the scores, which serve as $w$, gained by
winning the singles games of official tournaments.
We analyze networks
of tennis players, which are indicative of the VIP-club
phenomenon.  For each sex,
the score distribution obeys a power law with an exponential cutoff,
as the rank-score plots in \FIG\ref{fig:rank-score} indicate.
The fact that the probability density 
decreases in the score $w$ means that better players are scarcer.  The
players constitute undirected networks by doubles partnership, and the
edges are defined by two players pairing up in any of the doubles
tournaments in a year.
Our model cannot account for some aspects of the data, such
as the homophily in nationality and score-independent individuality in
the frequency of participation in the tournaments.  Nevertheless, two
players should be strong in total to be successful,
which provides thresholding. Simultaneously,
a strong player is expected to stick to a small
number of partners because of the paucity of similarly strong
players and the aversion to tying with weak
players. This is equivalent to homophily.

\begin{figure}
\begin{center}
\includegraphics[height=1.75in,width=2.25in]{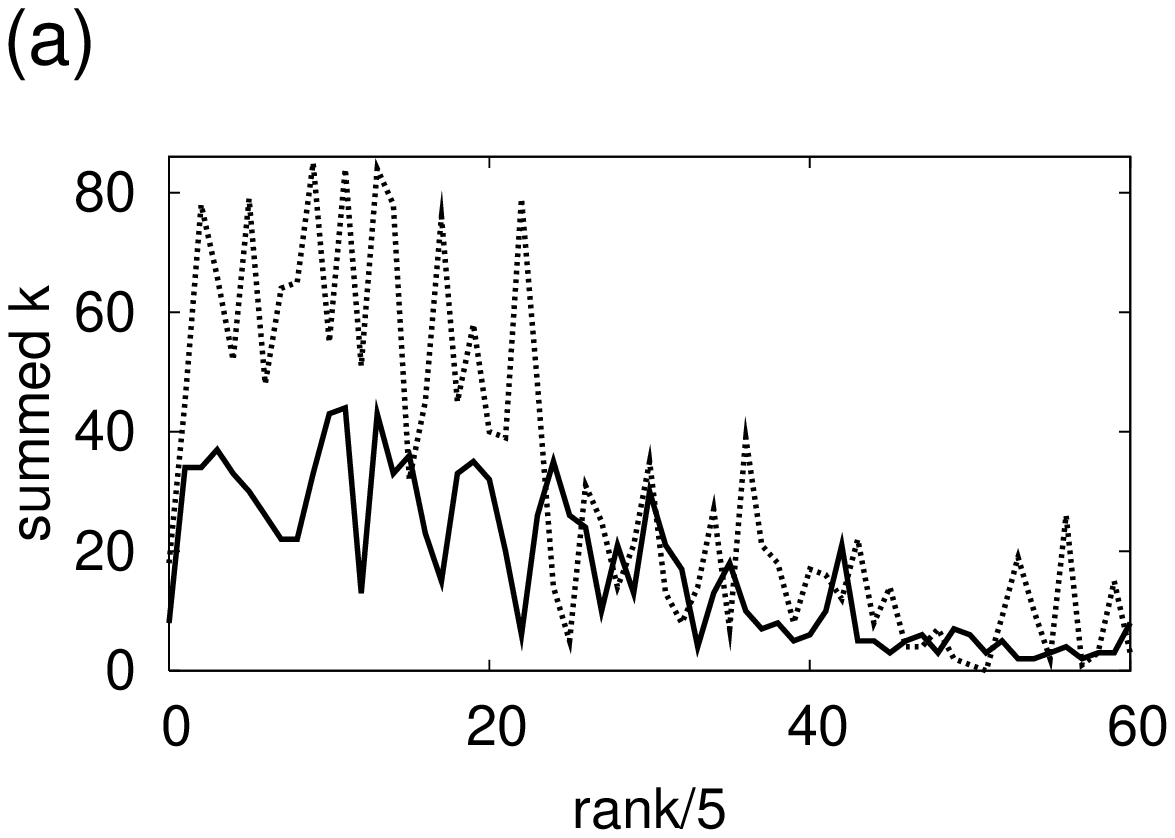}
\includegraphics[height=1.75in,width=2.25in]{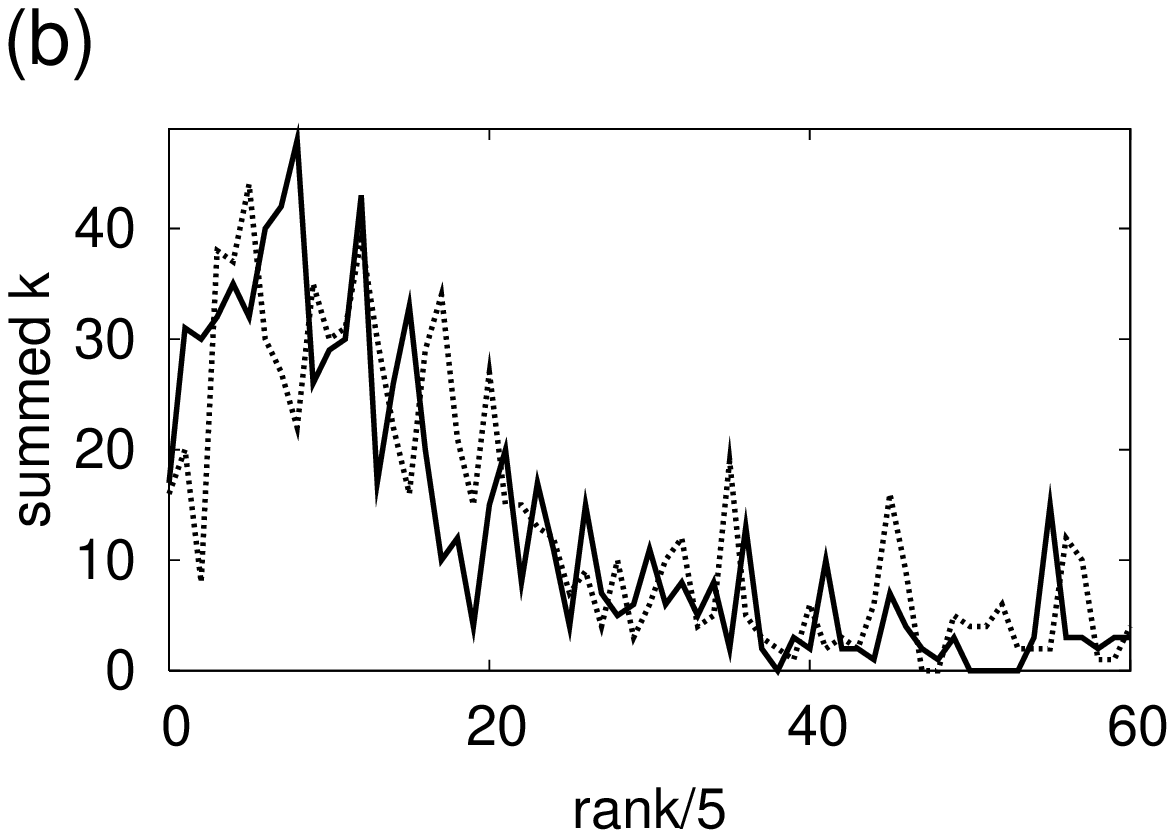}
\caption{}
\label{fig:tennis}
\end{center}
\end{figure}

The women's network based on all the WTA doubles tournaments in 2003
\footnote{http://www.wtatour.com/rankings/singles\_numeric.asp}
has $n=366$ excluding the isolated vertices.  Similarly, the
principal connected component of the men's network based on the ATP
doubles tournaments in 2003
\footnote{http://www.stevegtennis.com/} has $n=367$.  We do not consider
mixed doubles in which a man and a woman team up.  Since the ranking
is updated every week, players are differently aligned
according to the rankings at the ends of 2002 and of 2003.  To circumvent
noise in the data, we add $k$, or the number of doubles partners, of
five players with consecutive ranks.  The results are shown for women
and men in \FIGS\ref{fig:tennis}(a) and \ref{fig:tennis}(b),
respectively.  Although noise is yet large and the degree
distributions are not scale-free, players with intermediate ranks are
somewhat more capable of encountering partners.  The data in other
years also have this tendency (data not shown).
Players with high scores form a loose VIP club.
Players with intermediate ranks are popular among both stronger and weaker
players.

The VIP-club phenomenon is also expected in other social
situations.  For example, when choosing
a partner of the life,
one may set a threshold on social
statuses or incomes. Since they are graduated
quantities subject to homophily
\cite{Marsden,Mcpherson01,Mcpherson87},
maybe for people to communicate efficiently and live in comfort,
those in
upper-middle classes may be more promising in finding partners
than those too close to the top. Similarly,
scientific collaboration networks may exhibit the VIP-club phenomenon
particularly in the fields like mathematics where coauthorship is
rather strict. A group of strong researchers
generally 
publishes their work in nice journals (thresholding),
and researchers with
close abilities may flock together to coauthor (homophily). The age also shows
homophily \cite{Marsden,Mcpherson01,Mcpherson87},
and the population density, or $f(w)$, naturally decreases in the age,
or $w$. Consequently,
business or social activities that accompany
thresholding on ages
can result in VIP clubs formed by the old.
More generally, hierarchical
organizations of the pyramid type, such as companies and bureaucracies
\cite{Watts02SCI,Dodds}, are likely to
have VIP clubs.


\section{Conclusions}

We have shown that the combination of homophily and thresholding on graduated
vertex weights induces networks with
elites.  Loss of homophily leads to the rich-club phenomenon
\cite{ZhouMondragon}, while unrealistically many hubs emerge if
without thresholding. A VIP club is invisible to the majority of
vertices with small weights. They can even escape the eyes of network
analyzers unless the weight-driven edge direction in
addition to the
predefined edge direction, which are more readily obtained,
are actually inspected \cite{Anghel}. Actually, elites
and the majority of
vertices with small weights
remain undistinguished if based on vertex
properties such as $k$, the closeness
centrality, the reach centrality, the betweenness centrality,
or the local clustering coefficient \cite{Newman03SIAM,Scottbook}.
They
occupy structurally equivalent or similar locations of a network
\cite{Scottbook}.  Paradoxically, complete understanding 
of connectivity and
edge direction
is not necessarily sufficient for 
knowing a network \cite{Anghel}. To
understand the nature of a network, intrinsic properties of
individual vertices must be taken into account.

\newpage

Figure captions

\bigskip
\bigskip

Figure 1: Theoretically evaluated $k(w)$ (solid lines) and $k_2(w)$
(dotted lines) for the threshold graph supplemented by homophily.
We set $\lambda=1$, and (a) $(\theta, c) = (6, 5)$, (b)
$(6, 100)$, and (c) $(0, 5)$.  For clarity,
$k_2(w)$ is normalized so that the maximum is 1.
(d) $p(k)$ for $(\theta, c) =(6, 5)$ (crosses), $(\theta, c)= (6, 100)$
(squares), $(\theta, c) = (0, 5)$ (circles), and $p(k)\propto k^{-2}$
(line).

\bigskip

Figure 2: Numerically obtained
(a) $k(w)$ and (b) $p(k)$ for
the modified threshold graph. We set $n=50000$, $\lambda = 1$,
$\theta=6$, and
$(\beta_1, \beta_2) = (1.5, 0.5)$ (crosses), $(\beta_1,
\beta_2) = (1.5, 0)$ (squares), and $(\beta_1, \beta_2) = 
(0, 0.5)$ (circles).  The lines in (a) and (b) correspond to
$k(w)\propto
e^{\lambda w}$ and $p(k)\propto k^{-2}$, respectively,
which are the predictions by the threshold graph without homophily.
 
\bigskip

Figure 3: (a) The results for the model by Goh et al. (2001) with
$n=50000$ and $\alpha=0.5$.  (a) $k(w)$ and (b) $p(k)$ for $\beta_2 =
5$ (crosses) and $\beta_2=0$ (squares). The theoretical
estimates $k(w)\propto w^{-\alpha}$ and
$p(k)=k^{-(\alpha+1)/\alpha} = k^{-3}$ are indicated by lines.

\bigskip

Figure 4: Dependence of the score ($w$) on the rank of
female (crosses) and male (circles)
tennis players, based on the ranking at the
end of 2003.

\bigskip

Figure 5: Vertex degrees of the (a) women's and (b)
men's tennis networks in 2003.
The number of partners summed over five players
with consecutive ranks are plotted. The players are arranged according
to the ranking at the ends of 2002 (dotted lines) and of 2003 (solid
lines).

\newpage
\clearpage

\clearpage

\clearpage

\clearpage

\clearpage

\end{document}